\documentclass[11pt]{article}
 \usepackage{amssymb}
\newcommand{\bra}{\begin{array}}
\newcommand{\era}{\end{array}}
\newcommand{\beq}{\begin{equation}}
\newcommand{\eeq}{\end{equation}}
\newcommand{\beqar}{\begin{eqnarray}}
\newcommand{\eeqar}{\end{eqnarray}}

\def\BC{\bb C}
\def\_\BC{\bbi C}

\def\bZ {\bar{Z}}

\def\( {\left(}
\def\) {\right)}
\def\[ {\left[}
\def\] {\right]}
\def\no2 {{\textstyle{n\over 2}}}

\def\ra {{\rangle}}

\newcommand{\om}{\omega}
\newcommand{\Om}{\Omega}
\newcommand{\lam}{\lambda}
\newcommand{\si}{\sigma}

\newcommand{\eps}{\epsilon}

\newcommand{\te}{\theta}

\newcommand{\pa}{\partial}
\newcommand{\al}{\alpha}

\newcommand{\del}{\delta}

\newcommand{\ka}{\kappa}

\newcommand{\st}{\star}

\newcommand{\ov}{\over}

\newcommand{\lb}{\label}

\newcommand{\NP}[1]{ {\it Nucl.~Phys.} {\bf #1}}

\newcommand{\PR}[1]{ {\it Phys.~Rev.} {\bf #1}}
\newcommand{\PRL}[1]{ {\it Phys.~Rev.~Lett.} {\bf #1}}

\newcommand{\MPL}[1]{ {\it Mod.~Phys.~Lett.} {\bf #1}}

\newcommand{\JP}[1]{ {\it J.~Phys.} {\bf #1}:\  Math.~Gen.~}

\newcommand{\JMP}[1]{ {\it J. Math.~Phys.} {\bf #1}}

\begin{document}
\thispagestyle{empty}

\thispagestyle{empty}
\begin{flushright}
ucd-tpg/06-04\\
 hep-th/0611300
\end{flushright}
\vspace{0.5cm}

\begin{center}

{\Large \bf Thermodynamical Properties of Hall Systems\\}

\vspace{0.5cm}

{\bf Ahmed Jellal $^{a,b,}$\footnote{E-mail : ajellal@ictp.it -- jellal@ucd.ac.ma}}
and {\bf Youssef Khedif $^{b,}$\footnote{E-mail : youssef.khedif@gmail.com}}\\
\vspace{0.5cm}

{\em $^a$Center for Advanced Mathematical Sciences, 
College Hall,   
American University of Beirut,\\
P.O. Box 11-0236, Beirut, Lebanon }\\ [1em]

{\it  $^b$Theoretical Physics Group, Laboratory of Condensed Matter Physics,
Faculty of Sciences, \\ Choua\"ib Doukkali University,
P.O. Box 4056,
24000 El Jadida, Morocco}\\[1em]

\vspace{2cm}

{\bf Abstract}

\end{center}
\baselineskip=18pt
\medskip

We study quantum Hall effect within the framework
of a newly proposed approach, which captures the
principal results of some proposals.
This can be established by considering a system of particles
living on the non-commutative
plane in the presence
of an electromagnetic field and quantum statistical mechanically
investigate
 its basic features.
Solving the eigenvalue equation, we analytically derive the energy levels
 and the corresponding wavefunctions. These will be used, at low
 temperature and weak electric field,
to determine the thermodynamical potential $\Om^{\sf nc}$ and related physical
quantities.
Varying $\Om^{\sf nc}$
with respect to the non-commutativity parameter
${\te}$, we define a new function that
can be interpreted as a $\Om^{\sf nc}$ density.
Evaluating the particle number, we show that the Hall conductivity of the
 system is  ${\te}$-dependent.
This allows us to  make contact with
quantum Hall effect by
offering different interpretations.
We  study the high temperature regime and discuss the magnetism
of the system.
We finally show that at  $\te=2l_B^2$,
the  system is sharing some common features
with the Laughlin theory.

\newpage

\section{Introduction}

The quantum Hall effect (QHE)~\cite{prange} is a physical
phenomenon that appears  in two-dimensional electrons at low
temperature and in the presence of a strong uniform magnetic
field. It is an interesting subject, not only because of its
precise quantized plateaus of the Hall conductivity~\cite{kli},
but also of its relationship to different theories and
mathematical formalism.
 Non-commutative (NC)
geometry~\cite{connes} is one of the most successful tools that has been
 used to deal with the basic features of QHE. This is not surprising
 because this phenomena itself is of the non-commutative nature~\cite{sakita,jelnpb}. 
In fact, to glue the system  in the lowest Landau level,
the confining potential energy should be strong enough.
 Therefore particles can not jump to the next level because of
the large gap energy. Mathematically this effect can be traduced
by a non-commuting spatial variables.


There are several works
have been reported on the subject
by quantum mechanically studying spinless particles. Among them,
we quote the authors in~\cite{dj} who analyzed electrons in uniform external
magnetic and electric fields, which move in a plane whose coordinates
are non-commuting. They  tuned on
the non-commutativity parameter $\theta$ such that electrons
moving in the non-commutative coordinates are interpreted as either
leading to the fractional QHE or composite fermions in
the usual coordinates. In fact, these results will be
discussed in terms of our main idea that will
be analyzed in the present paper.

Motivated by the intrinsic relation between the NC geometry and
QHE, we develop an approach to investigate the basic features of
QHE. Based on the statistical mechanics technical, we analyze the
thermodynamical properties of particles on the NC plane  ${\mathbb
R}^2_{\te}$. In this analysis, we involve the  spin as an
additional degree of freedom. More precisely, we consider
particles of spin ${1\over 2}$
 living on ${\mathbb R}^2_{\te}$ in  the presence of an uniform electromagnetic
field and study its thermodynamical properties. Indeed, to
characterize the system  behaviour at low temperature as well as
weak electric field, we determine the thermodynamic potential
(TPO). This can be
 done by evaluating the corresponding physical quantities.
Getting  the particle number,  we derive the Hall conductivity
$\sigma_{\sf H}^{\sf nc}$ in terms of $\te$. It allows us to
define the widths of different plateaus
 entering
in the game as $\te$-function. Moreover by using some data, we
show that $\te$ can experimentally be fixed. Giving different
values to $\te$, we differently interpret $\sigma_{\sf H}^{\sf
nc}$. Taking
 advantage of the non-commutativity,  we obtain a general
 results those can be used to
explain fractional QHE, composites fermions and
 multilayer Hall systems.
We analyze with the
non-commutativity formalism the high temperature regime.
In fact,
the magnetism of our system will be discussed
by writing down the critical point. In the end, we show
that at the point $\te=2l_B^2$ our system behaves like
Laughlin one for the filling factor $\nu=1$.

The present paper is organized as follows. In section 2, the
Hamiltonian describing the system under consideration
on the ordinary plane  ${\mathbb R}^2$ will be given
as well as its spectrum.
This will be generalized by
adopting the star product and therefore will offer for us a mathematical tools
to deal with our proposal. In section 3,
basing on the
Fermi--Dirac statistics, we determine the corresponding partition function
that leads to TPO. This will be used to define particle number,
magnetization and a TPO density. In section 4, using the
standard definition,
we obtain a general form for the Hall conductivity. After getting the
filling factor, we offer different interpretations and establish
some links with other results related to QHE, in section 5.
The high temperature analysis and
the magnetism of the system will be considered in section 6.
Section 7 will devoted to discuss the critical point, i.e. $\te=2l_B^2$.
We conclude and give some perspectives in the last section.

\section{Hamiltonian and spectrum}

To deal with our task, we need to fix some mathematical tools. Firstly,
we define the Hamiltonian system on  ${\mathbb R}^2$ and
give its spectrum. Secondly, by adopting the NC geometry
we show how the obtained results on   ${\mathbb R}^2$  can be generalized
to those on ${\mathbb R}^2_{\te}$.

\subsection{Ordinary plane ${\mathbb R}^2$}

We consider one particle of spin ${1\over 2}$ and  mass $m_0$
living on ${\mathbb R}^2$  in the presence of
an magnetic ${\vec B}= B{\vec e}_z$
and electric ${\vec E}= E{\vec e}_y$ fields. Assuming that
this  system is confined in a finite surface such that $S=L_xL_y$.
In the Landau gauge
\beq
{\vec A}= (-yB,0, 0)
\eeq
the system is described by the Hamiltonian~\cite{landau}
\beq\label{HAM1}
H = {1\over 2m}\left( {p}_x -{eB\over c}
{y}\right)^{2} + { {p}_y^2\over 2m} + g\mu_{B}
\hat{S}B+{eE}{y}
\eeq
where $m$ is the effective particle mass in a crystal lattice. In (\ref{HAM1}),
 the third term is resulting from the interaction between spin
 and $\vec B$,  the last is reflecting the dipolar momentum and
 $\vec E$ interaction,  $g$ is the Land\'e factor and
$\mu_B={e\hbar\over 2m_0c}$ is the Bohr magneton.

The spectrum of $H$ can analytically be obtained by solving
the eigenvalue equation
\beq
H\Psi = E \Psi.
\eeq
 Consequently, the energy levels are
\beq\lb{els}
E_{n,m_s}^{{\tilde p}}=\hbar\om_c\left(n+ {1\over 2}\right)+g\mu_{B}
m_sB-{E\over B}c {\tilde p},\qquad  n=1,2,3\cdots,\
m_s=\pm {1\over 2}
\eeq
where $\om_c={eB\over mc}$ is the cyclotron frequency
 and  ${\tilde p}$ is a new momentum defined as
\beq
{\tilde p} = -p_x +{mcE\over 2B}.
\eeq
It  is submitted to the constraint~\cite{landau}
\beq\lb{ptil}
 \arrowvert\tilde p\arrowvert
\leq {eBL_y\over2c}
\eeq
for a weak electric field. The wavefunctions corresponding to (\ref{els}) are given by
\beq
\Psi_n(x,y) = e^{{i\over \hbar }{p}_x x }
\exp\left[ -{1\over 2l_B^2} \left( y - {y}_0 \right)^2 \right]
H_n \left[{1\over l_B}( {y} - {y}_0) \right]
\eeq
where $H_n$ are the Hermite polynomials and $l_B=\sqrt{\hbar c\over eB}$ is the
magnetic length. This latter plays an important role in the QHE world. In fact,
it defines the area occupied by the quantum Hall droplet
for particles in the lowest Landau level..

To close this part, we mention that the above results have been used
to deal with the
thermodynamical properties for the Hall particles of
 spin ${1\over 2}$ on  ${\mathbb R}^2$~\cite{hky}. More precisely,
integer QHE as well as multilayer Hall
systems have been discussed. Later, we propose a newly approach based on
the NC geometry that captures the basic features of~\cite{hky}
and leads  to different results.

\subsection{Non-commutative plane ${\mathbb R}^2_{\te}$}

As we claimed before, to deal with our task we may proceed by using the NC
geometry~\cite{connes}. This can be done by introducing the
commutator
\begin{equation}
\left[x_{j},x_{k}\right]=i\te_{jk} \label{nccoo}
\end{equation}
where
$\te_{jk}=\eps_{jk}\te$ is the non-commutativity parameter,
  $\te$ is a real and free quantity.
This relation can be realized by considering
the star product definition
\begin{equation}
\lb{defi} f(x) \st g(x)=\exp\left\{{i\over 2}\te_{jk}
\pa_{y^{j}}\pa_{z^{k}}\right\} f(y)g(z){\Big{|}}_{x=y=z} \label{2}
\end{equation}
where $f$ and $g$ are two arbitrary
functions, supposed to be infinitely differentiables. In what
follows, we will use the standard commutation relations
\begin{equation}
\lb{deqm}
[p_{j},x_{k}]=-i\hbar \del_{jk}, \qquad [p_{j},p_{k}]=0
\end{equation}
supplemented by the relation~(\ref{nccoo}), which together define a
generalized quantum mechanics and leads to the standard case once
$\te$ is switched off.

Now we show how the above tools can be employed
to define the non-commutative version of the Hamiltonian~(\ref{HAM1}).
In doing so, we start by noting that $H$ acts on an arbitrary function
$\Psi(\vec{r},t)$ as
\begin{equation}
\lb{hdef}
H \st \Psi (\vec{r},t) = H^{\sf nc}
\Psi (\vec{r},t).
\end{equation}
Applying the definition~(\ref{defi}), we obtain the required Hamiltonian
\beq \lb{HNC}
H^{\sf nc} = {1\over 2m}\left( \tilde{p}_x -{eB\over \al c}
\tilde{y}\right)^{2} + {{p}_y^2\over 2m} + g\mu_B
\hat{S}B+{eE\over \al^2}\tilde{y} -{mc^2E^2\over
  2B^2}\left(1-{1\over\al^2 } \right)
\eeq
where the quantities $\tilde{p}_x$ and $ \tilde{y}$
are given by
\beq
\tilde{p}_x =  \al {p}_x,\qquad \tilde{y} =  \al y + {mc^2E\over
  eB^2 } \left(\al -1\right).
\eeq
The important parameter of our theory is
\beq\lb{alpha}
\al= 1-{\te\over 2l_B^2}.
\eeq
We have some remarks in order.
 First, comparing~(\ref{HAM1}) with~(\ref{HNC}), one can see that~(\ref{HNC})
is including extra terms. Otherwise,
it is clear that $H^{\sf nc}$ is a generalized version of (\ref{HAM1}).
Therefore, it is  a good task to deeply  investigate the consequences of
$H^{\sf nc}$ on the Hall effect.
Second, according to (\ref{alpha}) one can notice that
there is a singularity at $\te=2l^2_B$ that should be
taking into account. For this, we offer
a section to discuss how, at this point,
the present system can be linked to the Laughlin theory.
Third,
without loss of generality,  we assume that $\al>0$ in the forthcoming
study.

To fully establish our mathematical tools, we need to derive
the spectrum of the non-commutative Hamiltonian~(\ref{HNC}). As usual,
this can be achieved
by solving the eigenvalue equation
\beq
H^{\sf nc} \Psi = E^{\sf nc} \Psi
\eeq
 to get the energy levels
\beq
E^{\sf nc}_{n,m_s,\tilde{p}_{\te} } = \hbar \tilde{\om}_c\left (n+{1\over 2} \right)
- c{E\over B} \tilde{p}_{\te} +m_s g{\hbar eB\over 2m_0c},
\qquad n=0,1,2\cdots, m_s =\pm{1\over 2}
\eeq
 where $\tilde{\om}_c$ and $\tilde{p}_{\te}$ are given by
\beq
  \tilde{\om}_c = {\om_c\over\al},\qquad
\tilde{p}_{\te}={-\al p_x }+{mc\over2}{E\over B}.
\eeq
In similar way to~(\ref{ptil}), for a weak electric field $\tilde{p}_{\te}$
 must fulfill the condition
\beq
 \left\arrowvert\tilde{p}_{\te}-\left(1-\al\right){mc\over2} {E\over B}
\right\arrowvert \leq\al {eBL_y\over2c}.
\eeq
As we will see
later, this will be helpful in order to approximatively evaluate
TPO on  ${\mathbb R}^2_{\te}$. The wavefunctions can be obtained
as \beq\lb{ncst} \Psi_n(x,\tilde{y},\te) = e^{{i\over
\hbar}\tilde{p}_x x } \exp\left[ -{1\over 2\tilde{l}_{B}^2} \left(
\tilde{y} - \tilde{y}_0
  \right)^2 \right]
H_n \left[ {1\over\tilde{l}_B}\left( \tilde{y} - \tilde{y}_0
  \right) \right]
\eeq
where the new quantity
\beq\lb{eeml}
\tilde{l}_B=\sqrt{\al \hbar c\over  e B}
\eeq
 can be interpreted as  an effective
magnetic length.
Moreover by analogy to $l_B^2$, (\ref{eeml}) can be seen as an effective
area of the quantum Hall droplet on  ${\mathbb R}^2_{\te}$.
For
$\al=1$, we recover the spectrum
 of the system on  ${\mathbb R}^2$ given before.

At this point, we have settled all formalism needed to do our
task. In fact, we will show it can be used
 to derive more general results those can be related to
the real physical systems. In doing so, let us begin by evaluating
different thermodynamical quantities.

\section{Thermodynamical properties}

In dealing with QHE, we start by applying quantum
statistical mechanics
 to derive different physical quantities. Indeed, we
evaluate the partition function of our system in order to
get the corresponding TPO. Subsequently, we introduce the
Mellin transformation
to obtain a simplified form of TPO. Consequently, we
 end up with two different residues, which showing
that TPO can be separated into two parts: monotonic and
oscillating PTO's. Combing all to derive the particle number and
related matters as well as discussing the magnetization.

\subsection{Thermodynamic potential}

Normally in statistical mechanics, TPO can be obtained
by determining the
partition function describing the system.
In our case, it is defined by
\beq\lb{z}
Z^{\sf nc} = {\rm Tr}  \left[\exp\left({\tilde\mu -H^{\sf nc}\over  k_BT}\right)\right]
\eeq
where
$\tilde\mu$ is the chemical
potential and trace is over all states given by~(\ref{ncst}).
Since we are considering a system of
fermions, the Fermi-Dirac statistics enforces to get 
\beq
Z^{\sf nc}=\prod_{n,m_s,\tilde{p}_{\te} } \left[1+\exp\left({\tilde{\mu}-
E^{\sf nc}_{n,m_s,\tilde{p}_{\te} }\over k_BT}\right)\right].
\eeq
This showing that TPO is
\beq\lb{tp1}
\Omega^{\sf nc}=-k_B T\sum_{n,m_s,\tilde{p}_{\te} }\log
\left[1+\exp\left({\tilde{\mu}-
E^{\sf nc}_{n,m_s,\tilde{p}_{\te} }\over k_BT}\right)\right].
\eeq
Keep in mind that $\tilde{p}_{\te}$ is a continue variable. By
introducing a set of the dimensionless
quantities
\beq
 p_{\te}={\tilde {p_
    {\te}}\over{mc}},\qquad\  {\mu}={{\tilde{\mu}}\over {mc^2}}, \qquad\
\lam ={k_BT\over mc^2}, \qquad\ \epsilon^{\sf
nc}_{n,m_s,\tilde{p}_{\te}} ={E^{\sf nc}_{n,m_s,\tilde{p}_{\te}
}\over mc^2}
\eeq
we write $\Omega^{\sf nc}$ as \beq\lb{tp2}
\Omega^{\sf nc} =- mc^2 \lam\al
 N_{\phi} \ \int_{{a-b\over 2}}^{{a+b\over 2}} dp_{\te} \
{1\over b} \ \sum_{n,m_s} \ \ln \left[ 1+\exp\left({\mu -\
\epsilon_{n,m_s,\tilde{p}_{\te} }^{\sf nc}\over \lam}\right)
\right] \eeq where $a$ and $b$ are given by \beq
a=\left(1-\al\right) {E\over B}, \qquad b= \al{eBL_y\over{mc^2}}.
\eeq
The number  $N_{\phi}$ is  the quantized flux:
 \beq
N_{\phi}={\phi\over\phi_0 }={eBS\over hc}.
\eeq
In the QHE world,
it is well-known that $N_{\phi}$ is an important  quantity that
should be determined. This comes from the fact that $N_{\phi}$ is
exactly the degeneracy of Landau levels, which defines the filling
factor, see subsection (4.1).

To calculate (\ref{tp2}),
we need to introduce a
relevant method. This can be realized by
making use of the Mellin transformation. In general, it is defined by
\beq\lb{mtra}
f(x) ={1\over  2\pi i}\int_{c-i\infty}^{c+i\infty}{\tilde
  f}(t)x^{-t}dt \ \ \ \Longleftrightarrow \ \ \
{\tilde  f}(t) =\int_{0}^{\infty} f(x)x^{t-1}dx
\eeq
for an arbitrary function $f(x)$. By
applying (\ref{mtra}) to our case
with respect to the  variable
$e^{{\mu/\lam}}$ and after a straightforward calculation, we
find
\beq\lb{omnc}
\Omega^ {\sf nc}= \mp mc^2
\lam\al N_{\phi} \sum_{s}{\mbox Res}\  {\pi \exp\left({s\mu\over\lam}\right) \over
  s\sin\pi s} \ \int_{{a-b\over 2}}^{{a+b\over 2}} dp_{\te}\ {1\over
b} \sum_{n,m_s}\ \exp \left( -{sE_{n}^{\sf nc}\over
mc^2\lam}\right)
\eeq
where the upper and lower sign refer,
respectively, to closing the contour to the left and right of the
imaginary axis for $\bar\mu>0$ and $\bar\mu<0$. The integral gives
\beq\lb{tp3}
\Omega^ {\sf nc}=\mp mc^2 \lam\al N_{\phi}
\sum_{s}{\mbox Res}\ \left[
    {\pi\exp^{s\bar{\mu}\over\lam}\over s\sin(\pi s)}\ {\cosh({
        s{g^\star}\kappa\over 2\lam})\over\sinh({s\kappa\over
        2\al\lam})}\ {\sinh({se\al EL_{y}\over 2\lam mc^2})\over
      {se\al EL_{y}\over 2\lam mc^2}}\right]
\eeq
and the new involved parameters
\beq
{g^\star}={gm\over 2m_0},\qquad \kappa={B\over {B^\prime}},\qquad
{B^\prime}={m^2c^3\over e\hbar},\qquad
\bar\mu=\mu+{a\over 2}{E\over B}.
\eeq
In conclusion,
 one has to distinguish two different cases. Firstly
$\bar\mu>0$,
the sum of residues can be calculated at the
negative poles in the real semiaxis $ s=0,-1,-2,-3,\cdots $  and in the
imaginary semiaxis at the poles
$s_l ={2\pi i l}{\al\lam\over\kappa}$. Secondly
 $\bar\mu<0$, the sum of residues can be calculated by closing the
integration contour to the right, i.e. considering only the positive real
semiaxis, however the residues can be evaluated at the poles
$s=1,2,3,\cdots$.

 It is not easy to directly evaluate $\Omega^ {\sf nc}$ in
(\ref{tp3}),
but it can be achieved by introducing
 some approximations.
For this, we consider
the high and low temperature regimes as
 well as assume that the electric field is weak. In doing so, we separate TPO
 into two parts: monotonic and oscillating.

\subsection{Low temperature regime}

To completely determine  TPO, we consider two different cases those are
 low and high temperature regimes. Note that,
the first case is governed by the constraint
$\lam \ll\bar{\mu}$. Since we have two poles, one can separate TPO into two
parts, which are the monotonic $\Omega_{\sf mo}^{\sf nc}$ and
oscillating $\Omega_{\sf os}^{\sf nc}$ terms corresponding, respectively,
to  two poles $s=0$ and $s_l ={2\pi i l}{\al\lam\over\kappa}$.

Now we evaluate each part separately. Indeed,
using the residue methods, one can show that
the monotonic part is
\beq\lb{tpmon}
\Omega_{\sf mo}^{\sf nc}=-mc^2 N_{\phi}\kappa \al
\left[{1\over4\pi^2}\al\xi^2 +{1\over4}\left({\al g^\star}^2 -
{1\over3\al}\right)+
{\pi^2\over3}\al\left(\lam\over\kappa\right)^2+\al^3{\eta^2\over12} \right]
\eeq
where $\xi$ and $\eta$ are
\beq
\xi={2\pi\bar{\mu}\over \kappa}, \qquad \eta={eEL_{y}\over\kappa mc^2}. \
\eeq
Note that, in deriving  $\Omega_{\sf os}^{\sf nc}$ we have dropped the contributions
resulting from other residues $s\neq 0$, because they are exponentially
small. On the other hand,
the oscillating TPO can
be evaluated by assuming that the condition $\lam \ll {\kappa\over \al}$
is holding. This implies that  $\Omega_{\sf os}^{\sf nc}$ can be written as
\beq\lb{tposc}
\Omega_{\sf os}^{\sf nc}=-mc^2 N_{\phi}\kappa\ \sum_{l=1}^{\infty}\ {(-1)^{l+1}}\
    {\cos(\al \xi l)\cos(\al {g^\star}\pi l) \over\pi^2 l^2}\
    {\sin(\pi\eta\al l)\over \pi\eta\al l}.
\eeq
 Therefore, combining all we get
\beq
\Omega^{\sf nc}_{\sf lo} = \Omega_{\sf mo}^{\sf nc} +\Omega_{\sf os}^{\sf nc}.
\eeq

As claimed before, we are considering a very weak electric field. This
assumption is equivalent to the constraint
\beq
 E \ll {\kappa
  mc^2\over eL_y} \ \ \Longleftrightarrow \ \ \eta \ll 1.
\eeq
In this limit, we show
\beq
\Omega^{\sf nc}_{\sf lo}=-mc^2 N_{\phi}\kappa \al \left[{1\over4\pi^2}\al\xi^2
+{1\over4}\left({\al g^\star}^2 -
{1\over3\al}\right)+ {\pi^2\over3}\al\left(\lam\over\kappa\right)^2
+{1\over\al\pi^2}F^{\sf
  nc} ({\xi})\right]
\eeq
where  $F^{\sf nc} ({\xi})$
is given by \beq\lb{ffunc} F^{\sf
nc} (\xi)= \sum_{l=1}^{\infty}\ (-1)^ {l+1}\ {1\over l^2}\
\cos(\al \xi l)\ \cos(\pi\al{g^\star}l). \eeq
It can be rearranged
as
\beq
F^{\sf nc} ({\xi})={1\over 2}\ \sum_{l=1}^{\infty}\ (-1)^
{l+1}\ {1\over l^2}\ [\cos\left\{\al\left(
\xi+{g}^\star\pi\right)l\right\} +\cos\left\{\al\left(
\xi-{g}^\star\pi\right)l\right\}].
\eeq
This function will play an
important role in discussing QHE on ${\mathbb R}^2_{\te}$. In
fact, we will see how it can be linked to the filling factor.

Statistical mechanics teaches us that the thermodynamical
potential is the first step that should be determined in order to
study the system behaviour. This is because of its relationships
to other physical quantities. To clarify this point, we start by
evaluating
the particles number:
 \beq\lb{ncnp} N^{\sf nc}_{\sf lo}=- {1\over
mc^2} {\pa{\Omega^{\sf
      nc}}\over\pa\bar{\mu}}.
\eeq
After a straightforward calculation, we find
\beq
N^{\sf nc}_{\sf lo}=N_{\phi}{\al\over\pi}\left[\al\xi +{2\over\al}
{\pa {F^ {\sf nc}(\xi)}\over\pa\xi}\right]
\eeq
which is
depending to $N_{\phi}$ and $\al$. In fact, this will be helpful
in defining the Hall conductivity on ${\mathbb R}^2_{\te}$ and
thus getting its final form.

 It might be a good task to see how
the magnetization of the present system does look like.
In doing so, we start by noticing that
the magnetic field is along $z$-direction, then
there is only one
component following the $z$ axis of the magnetic moment of particles
on $\mathbb{R}^2_{\te}$. It is given by
\beq
M^{\sf nc}=- {\pa{\Omega^{\sf nc}}\over\pa {B}}.
\eeq
 It can be calculated to get
\begin{eqnarray}
\lefteqn{
M^{\sf nc}_{\sf lo}  = mc \bar{\mu} {eS\over\hbar}
\Biggl[ {\xi\over 2\pi^2} \left(\al^2-\al\right)+ {1\over
    2\xi}\left(\al\left(2\al- 1\right){g^\star}^2-{1\over
    3}\right)
+{2F^{\sf nc}(\xi)\over\pi^2\xi}
~{} } \nonumber \\
& &
~~
+{2\pi^2\over
    3\xi}\left(\al^2-\al\right)\left(\lam\over\kappa\right)^2
-{1\over \pi^2}{\pa{F^{\sf nc}
    (\xi)}\over\pa{\xi}}-{a\over \pi\kappa } {E\over B}
\left({\al^2\over 2}+{1\over \xi}{\pa{F^{\sf nc} (\xi)}\over \pa{\xi}}
\right)
\Biggr].
\end{eqnarray}
The obtained form of $M^{\sf nc}_{\sf lo} $ can deeply
be investigated to make contact
with early works on the magnetization~\cite{ich,gaz,jel1}.
With this, one may have a full picture on the magnetism
of the system at low temperature.

We are not going to stop at this level in
deriving physical quantities. In fact,
since TPO is $\te$-dependent, one may define a new function
by making variation with respect to $\te$. This is equivalent to write
\beq
\chi^{\sf nc}_{\sf lo}= -{\pa{\Omega^{\sf nc}_{\sf lo}}\over\pa{\te}}
\eeq
which leads to
\beq\lb{xi}
\chi^{\sf nc}_{\sf lo}=mc \bar{ \mu} {eS\over\hbar}\left(\al-1\right)\Bigg[{\al\over
    2\pi^2}\xi
- {\al^2\over 2\pi\kappa }\left({E\over B}\right)^2
+{\al{g^\star}^2\over 2\xi}
+{2\pi^2\over 3}{\al\over \xi}\left({\lam\over \kappa}\right)^2
+{1\over\pi^2\xi} {\pa{F^{\sf nc}}
(\xi)\over\pa{\al}}
\Bigg].
\eeq
Because of $\te$ has length
square of dimension, thus $\chi^{\sf nc}_{\sf lo}$ can be interpreted as a TPO density of a
system possessing $\te$ as surface. This is the case for instance for a quantum
Hall droplet where its area is $l_{B}^{2}={\hbar c\over eB}$, which
corresponds to $\al={1\over 2}$ in our case. Of course
(\ref{xi}) goes to zero once we set $\al=1$.

\section{Hall conductivity}

To make contact with the Hall effect, we should determine the
Hall conductivity in terms of our language. It can be done
by adopting the usual definition of the Hall current
as well as using the Heisenberg equation for the particle
velocity on $\mathbb{R}^2$.
As we will see soon, these will offer for us a
 general form of the
filling factor that can be used to
show that our system is describing a real physics.

\subsection{Filling factor on $\mathbb{R}^2$}

To establish different definitions to deal with
our issues on
the generalized plane,
let us sketch some discussion about the Hall conductivity  on
$\mathbb{R}^2$. We start from classical to quantum mechanics in order
to emphasis the differences.

Let us consider
system of $N$ particles living on the plane
in the presence of magnetic  $\vec{B}=B \vec{e}_z$ and
electric  $\vec{E}=B \vec{e}_y$ fields.
Experimentally,
 the classical
result is showing that the inverse of $\sigma_{\sf H}$ is behaving
like a straightforward line in terms of $B$, which has given birth
to the Hall effect.
Theoretically, this
effect can
be explained by studying Newton  mechanically
the  motion of a Hall system. In the end, one can obtain
the Hall conductivity
\beq\lb{HC1}
\sigma_{\sf H}=-{\rho e c\over B}
\eeq
where the particle density $\rho={N\over S}$ is resulting
from the Hall current definition.
This is showing how the conductivity is linked to the magnetic field.
Moreover, it is easily seen that (\ref{HC1})
 is actually  reflecting
the experiment result.

Now let us see how  $\sigma_{\sf H}$ does look like
in quantum mechanics. Resorting the spectrum of the Hamiltonian
describing the above system, without interactions, we can
derive
$\sigma_{\sf H}$
in terms of the fundamental constants. This is
\beq\lb{HC2}
\sigma_{\sf H}=-\nu {e^2\over h}
\eeq
where the filling factor $\nu$ is given by
\beq\lb{FF}
\nu={\rho ch\over eB}=2\pi l_B^2 \rho.
\eeq
Clearly, $\nu$ is  actually measuring different values of $B$
for a fixed density $\rho$.
Furthermore, to talk about QHE $\nu$~should be
quantized over a large range of a strong magnetic.
This can be traduced by defining  $\nu$ as
\beq\lb{qhe}
\nu ={N\over N_{\phi}}
\eeq
where $ N_{\phi}$ represents also the degree of degeneracy
of each Landau level. Otherwise, $\nu$ is the ratio between
particle number and  number of the accessible states
per Landau level. To get integer QHE or fractional QHE,  $\nu$
should be, respectively, integrally or fractionally quantized.

In conclusion, we need to determine the analogue of $\nu$
on $\mathbb{R}^2_{\te}$ and see how can be used
to offer different interpretations and establish
some links. To do this end, we assume that all definitions
above will be also valid on the non-commutative plane.

\subsection{Effective filling factor}

According to the above analysis, it follows that the filling
factor is a relevant quantity, which has to be determined
in having the Hall effect. In doing so, we use
(\ref{HC2})
to write the non-commutative
Hall conductivity as
\beq
\sigma_{\sf H}^ {\sf nc}=-\nu^{ \sf nc}
      {e^2\over h}.
\eeq
where $\nu^{ \sf nc}$ is its corresponding effective filling factor.
Using  (\ref {qhe}),  $\nu^{ \sf nc}$ can be seen as ratio between the particles
 number $N^{\sf nc}_{\sf lo}(\xi)$ and the degeneracy of the Landau levels
 $N_{\phi}$, i.e. quantum flux. Otherwise, one may write  $\nu^{ \sf nc}$ as
\beq\lb{edf}
\nu^{\sf nc}(\xi)={N^{\sf nc}_{\sf lo}(\xi)\over N_{\phi}}.
\eeq
Since we have all ingredients, it is easy to see
\beq\lb{fornu}
\nu^{\sf nc}(\xi)={\al\over\pi}\left[\al\xi+{2\over\al}{\pa F^{\sf nc}(\xi
  )\over\pa{\xi}}\right].
\eeq

Note that, (\ref{fornu}) is showing that there is a relation
between the non-commutativity of space and the filling factor of
the present system. To clarify this point, we simplify $\nu^{\sf
nc}(\xi)$ to another form. Indeed, using the identity
\beq
\sum_
{l=1}^ {\infty}\ (-1)^{l+1}\ {\cos(lx)\over l^2}={\pi^2\over
  12}-{x^2\over 4}
\eeq
we show that $F^{\sf nc}(\xi)$ can be written as
\beq\lb{ffun}
F^{ \sf nc}(\xi)=\left\{ \begin{array}{ll}
{\pi^2\over 12} - { \al^2\over 4}{\xi}^2 - {
  1\over4}{\al^2{g^\star}^2\pi^2} \
\qquad\qquad\qquad\qquad\qquad\qquad \textrm{if }\  -u_1\leq  \xi\leq u_1
\\
 {\pi^2\over 12} - { \al^2\over 4}{\xi}^2 - { 1\over4}{\al^2{g^\star}^2\pi^2}
 +\al\pi\xi-{1\over2}\left(1-\al{g^\star}\right)\pi^2   \, \qquad
\textrm{if } \ \  \  \  \, u_1\leq \xi \leq u_2
\end{array} \right.
\eeq
where $u_1$ and $u_2$ are   given by
\beq
u_1=\left ({1\over\al}-{g^\star}\right)\pi,
\qquad u_2=\left({1\over\al}+{g^\star}\right)\pi.
\eeq
Moreover,
 (\ref{ffun}) can be generalized to any odd integer value $i$, such as
\beq
F^{ \sf nc}(\xi)=\left\{ \begin{array}{ll}
{\pi^2\over12} -  {1\over4}\al^2 {\xi}^2 - {1\over4} {
    g^\star}^2\al^2\pi^2 - {\pi^2\over4} -
{1\over4}i^2\pi^2 + { 1\over2}i\al\xi\pi +  {1\over2}\al{g^\star }\pi^2
 \,\,\,  \qquad \textrm{if }\
u'_1\leq \xi \leq u'_2
\\
{\pi^2\over12} -  {1\over 4}\al^2 { \xi}^2 - { 1\over4}{
    g^\star}^2\al^2\pi^2 - i^2 { \pi^2\over 4} - {\pi^2\over 4} -
 {1\over 2}i\pi^2 + \al\xi (i+1) {\pi\over2}
 \  \qquad \textrm{if }\
u'_2\leq \xi \leq u'_3
\end{array} \right.
\eeq
where $u'_1$, $u'_2$ and  $u'_3$ are
\beq
u'_1= \left({i\over\al }-{g^\star}\right)\pi,\qquad u'_2=
\left({i\over\al}+{g^\star}\right)\pi,
\qquad u'_3= \left({i+2\over\al}-{g^\star}\right)\pi.
\eeq
Therefore, the final form of the filling factor is given by
\beq\lb{nff}
\nu^{ \sf nc}(\xi)=\left\{ \begin{array}{ll}
0 \ \qquad\qquad\qquad  \textrm{if } \  \ 0\leq \xi \leq
u'_1\\
i\al  \ \ \,\quad\qquad\qquad \textrm{if } \  \ u'_1\leq \xi \leq
u'_2\\
  (i+1)\al\ \ \ \qquad
\ \textrm{if } \ \ u'_2\leq \xi\leq
u'_3.
\end{array} \right.
\eeq
 This  is more suggestive and thus one needs
 to look deeply for its interpretation. Moreover, we will see that
the different intervals can be used to discuss a possible
measurement of $\te$.
These materials will be the subject of the next section.

We still need different definitions settled above for the
Hall effect on $\mathbb{R}^2$. Indeed,
 by comparing (\ref{FF}) with  (\ref{nff}), it follows that one can
define an effective
magnetic field as
\beq\lb{nmf}
B_{\sf eff}={B\over \al}\equiv{B\over1-{\te\over 2l_B^2}}.
\eeq
This implies that its effective filling factor is
\beq\lb{rff}
\nu_{\sf eff} \equiv\nu^{ \sf nc}(\xi) = {\rho c h\over eB_{\sf eff}}
\eeq
where the quantity  ${\rho c h\over eB}$ is integrally quantized
and more specifically is corresponding to the case when $\al=1$ in~(\ref{nff}),
namely $\nu^{ \sf nc}(\xi)|_{\al=1}$.

Equations (\ref{nmf}) and (\ref{rff}) are interesting results in sense that
they will
be employed to clarify our motivation
behind the present work. In fact, we use $B_{\sf eff}$
and $\nu_{ \sf eff}$ to offer some interpretations
and also show different bridges between our proposal and some
results related to QHE.

\section{Interpretation} 

In the beginning, let us notice that
particles of spin ${1\over 2}$ living on $\mathbb{R}^2_{\te}$ when
uniform external magnetic as well as electric fields are
applied can be envisaged as the usual motion of particles
experiencing an effective magnetic field (\ref{nmf}). This
can be regarded as standard results, but let us go deeply
to see
what we can do with different values of $\te$.

\subsection{Measuring $\te$}

It is  a good job to look for an experiment realization of  the
non-commutativity parameter  $\te$. For this, we offer one
possibility that can be used to identify our system to {\it GaAs}.
In doing so, let us return to comment the filling
factor~(\ref{nff}).

It is clear that the widths of the plateaus coincide with
different intervals appearing in different values of (\ref{nff}).
For the second and third equations in~(\ref{nff}), the widths,
respectively, are given by \beq\lb{wid} W_{21} = 2\pi g^*, \qquad
W_{32} = 2\pi \left({1\over \al}-g^*\right). \eeq Consequently,
there are two critical points those should be discussed in order
to get more information about our system. These are \beq
g_{21}^*=0, \qquad g_{32}^*= {1\over \al}. \eeq Let us discuss
each case separately. Indeed, $ g_{21}^*$ is actually
corresponding to the filling factor \beq\lb{camdj} \nu^{ \sf
nc}(\xi)=  (i+1) \al. \eeq It follows that, there is no splitting
of spin degree of freedom in each Landau level. However, in the
second case all Landau levels except the ground state are spin
degenerate and we have \beq \nu^{ \sf nc}(\xi)=  i \al. \eeq It
turns out that $ g_{32}^*$ is important value because it is linked
to the non-commutativity of the system. In fact, since $\te$ is a
free parameter one can get different $ g_{32}^*$ in order to link
our system to different real physics systems. In what follows, we
offer an experiment measurement of $\te$ by using some data.
Indeed, according to~\cite{stone} for the Hall system {\it GaAs},
we have \beq\lb{edata} m \approx 0.07 \, m_0, \qquad g\approx 0.8,
\qquad g^* = 0.03. \eeq This implies that at the second  point,
one may fix  $\te$ as \beq \al \approx 33.33 \ \
\Longleftrightarrow \ \ \te \approx -64.66 \, l_B^2. \eeq
Thus, when  $\te \approx -64.66 \, l_B^2$ our system can be regarded as
 {\it GaAs}. In conclusion,
this results is interesting in sense that one may use different experiment data
to see how the non-commutativity parameter can experimentally be detected.
Note that, one possibility has been
realized by one of the present authors in~\cite{manar}.

\subsection{Dayi and Jellal approach}

Dayi and Jellal proposed an approach based on the NC geometry
tools to describe the Hall effect of a system of
electrons~\cite{dj}. In fact, their corresponding filling factor
is found to be
\beq \lb{dj} \nu_{\rm dj} = {\pi\ov 2} \rho (
l_B^{'2} - \te^{'} )
\eeq
which is identified to the observed
fractional values
for FQHE. This
approach also allowed to make a link with
the composite fermion approach
\cite{cf}
 by setting an effective
magnetic field
\beq\lb{djeff}
B_{\rm dj} = {B\ov 1-\te' l_B^{'-2}}
\eeq
similar to that felt by the composite fermions, with  $ l_B^{'}=2l_B$
and $\te'$ is a non-commutativity parameter.

Now let see if  the above approach has some overlapping
with our analysis. We start by noting that
the relations (\ref{dj}) and (\ref{djeff}) have been
obtained for a spinless electrons. The fact that of taking into
account of spin
as an additional degree of freedom plus
the system confinement in rectangular area  offered for us a possibility
to have plateaus 
and therefore
an experiment measurement for $\te$. On the other hand,
for zero spin, one may have the identification
\beq
\nu_{ \sf eff}|_{\te=\te_{\sf dj}}=\nu_{\rm dj}
\eeq
at the critical point $g^*=0$ and thus the comparison should be done
with respect to (\ref{camdj}). It follows that, one can solve the
 equation
\beq\lb{rlfj}
(i+1)\al_{\sf dj} = {\pi\over 2} \rho l_B^{'2} \al'
\eeq
to establish different relations between $\te$ and $\te'$, or
let say between $\al$   and $\al'$ with
\beq \al'= 1-{\te'\over
l_B^{'2}}.
\eeq
It is easily seen that one may have
\beq
\al_{\sf dj} ={\nu\over 2(i+1)} \al'
\eeq
where the filling factor $\nu$ is given by (\ref{FF}). In terms of
$\te$, we obtain
\beq
\te_{\sf dj} = 2l_B^2 - {\nu\over 2(i+1)} (2l_B^2 -\te').
\eeq
It reflects that the results derived in~\cite{dj} can be
linked to the present approach if our parameter $\te$ is fixed to be $\te_{\sf dj}$.
It is obvious that, whenever we have $\te=\te'$,
$\nu$ should be quantized as even integral values, such as
\beq
 \nu = 2(i+1).
\eeq
This is showing that there is missing part of the filling
factor, which is the odd values. Therefore, we have a partially
integer QHE. To overcome this situation, we will show how to get a
full picture of the integer QHE.

\subsection{Other approaches}

We are looking for other explanations of our results. Indeed, by
tuning on  $\te$ we  offer four different interpretations of
$\nu_{ \sf eff}$. These are related to integer QHE, fractional
QHE, composite fermions and multilayer Hall systems.

$\bullet$ {\sf Integer quantum Hall effect}: it is
corresponding to an integral quantized Hall conductivity, i.e.
the filling factor is integer. To reproduce this effect from our
consideration, we simply set
\beq
\al=1
\eeq
which basically corresponds to $\te$ is switched off.
In this case our analysis coincide with
that has been reported on the subject in~\cite{hky}. Consequently,
this  result is showing that our
approach is relevant for QHE and
is general in sense that one can get other links.

$\bullet$ {\sf Fractional quantum Hall effect}: is one of the most interesting
features of low-dimensional systems~\cite{qhe}.
It is characterized by the observed
Hall conductivity
\beq
\si_{\sf H}=f \;{e^2\over h}
\eeq
where $f$ is denoting the fractional quantized values of the
filling factor $\nu .$
This effect can be interpreted
in terms of the Hall effect on $\mathbb{R}^2_{\te}$.
More precisely, we identify the
effective filling factor (\ref{rff}) with the observed value $f$
by requiring that
\beq
\lb{fff}
\nu_{\sf eff}|_{\te=\te_{\sf f}}=f.
\eeq
This can be solved by considering different plateau widths.
According to (\ref{wid}), for $W_{21}$ we get the first critical value
of $\te$
\beq
\lb{fff1}
{\te_{\sf f_1}} = 2l_{B}^{2}\left(1-{f\over i}\ \right).
\eeq
By considering  $W_{32}$, we obtain the second
\beq
\lb{fff2}
{\te_{\sf f_2}} = 2l_{B}^{2}\left(1-{f\over i+1}\ \right).
\eeq
Therefore, when $\te$ is fixed to be $\te_{\sf f_i}$, with $(i=1,2)$,
one can envisage the Hall effect on
$\mathbb{R}^2_{\te}$ as the usual fractional QHE on the plane.

$\bullet$ {\sf Composite fermions}: are new kind of particles appeared in
condensed
matter physics to provide an explanation of the behavior of particles
moving in the plane when a strong magnetic field $B$
 is present~\cite{cf}. Particles
possessing $2p$, with $p\in N^*$, flux quanta (vortices) can be
thought of being composite fermions. One of the most important features of
them is they feel effectively a magnetic field given by
\beq
\lb{cfm}
B^*=B - 2p\Phi_0\rho.
\eeq

To interpret particles of spin ${1\over 2}$ living on $\mathbb{R}^2_{\te}$
in the magnetic field $B$ and electric field $E$
as the usual composite fermions we should tune on $\te$, such as
\beq
\lb{emf}
B_{\sf eff}|_{\te=\te_{\sf cf}}=B^{*}
\eeq
where $B_{\sf eff}$ is given by~(\ref{nmf}).
We solve~(\ref{emf}) to obtain
\beq
\lb{cte}
\te_{\sf cf}=2l_{B}^{2}\left[1- \left(1-{2p\phi_0\rho\over
    B}\right)^{-1}\right].
\eeq
In the limit of a strong magnetic field, it leads
\beq
\lb{ctepm}
\te_{\sf cf}\approx-4p\pi\rho l_{B}^{4}.
\eeq
Therefore, composite fermions can be envisaged
as particles living on  $\mathbb{R}^2_{\te}$ in the magnetic field $B$ and
the electric field $E$ when we set
$\te =\te_{\sf cf}$.

$\bullet$ {\sf Multilayers}:
structure formed by several identical Hall systems and its integral
quantization of the total Hall conductivity is given by~\cite{tsui}
\beq
\sigma_{\sf H}^{\sf tot}=J\sigma_{\sf H}^{i}
\eeq
where $J$ is the number of the layers including in such structure and
$\sigma_{\sf H}^{i} $ corresponding the Hall conductivity in a  single
layer.

To describe the integer quantization of multilayers from
the motion of particles of spin ${1\over 2}$ on $\mathbb{R}^2_{\te}$  in the presence of
$B$ and $E$, one may quantize $\al$ as
\beq
\al_{\sf J}=J.
\eeq
This is reflecting that how this kind of Hall
structures  can be identified to our system
living on the non-commutative plane.

In conclusion, by tuning $\te$ on we got different interpretations.
These concern integer QHE, fractional QHE, CF as well as multylayer Hall systems.
Therefore, the above results are proving that our analysis is relevant
for the Hall effect.

\section{High temperature analysis}

To complete our analysis, we need to discuss the high temperature regime
of the present system on
the non-commutative plane. Mathematically, this regime is
equivalent to have the limit $\lam  \gg \kappa$. Moreover,
our discussion will be  achieved by requiring that
the conditions
$\eta \ll1$ and $\bar {\mu}< 0$ are satisfied.
These will be used to get the corresponding TPO and related matters. In particular,
we analyze the magnetism by showing that there is a phase transition.

In getting TPO at high temperature, we evaluate~(\ref{tp3})
by closing the contour to the right of the imaginary axis and taking
the lower sign. This is
\beq
\Omega^{\sf nc}_{\sf hi}=mc^2\lam N_{\phi}\ \al\sum_{s}\ \mbox Res\  {\pi e^
  {s\bar {\mu}\over \lam}\over s\sin \pi s}\ {\cosh \left(
  {s{g^\star}\kappa\over2\lam}\right)\over \sinh
\left( {s\kappa\over2\al\lam}\right)}.
\eeq
Under the previous conditions, we find
\beq\lb{htp}
\Omega^{\sf nc}_{\sf hi}\cong mc^2\lam N_{\phi}\ \left[ 2\al^2 {\lam\over
    \kappa} \ {Li_2}\left(-e^ {\bar\mu/ \lam}\right)\ -
  \left( {\al^2 g^\star}^2-{ 1\over 3}\right)
{\ka\over 4\lam\left(1+e^{-\bar{\mu}/ \lam}\right)}\right]
\eeq
up to the first order of ${\kappa\over\lam}$. This is describing the system behaviour
at high temperature, which  goes to its analogue on $\mathbb{R}^2$ by choosing
 $\al=1.$

Similarly to low temperature case, we use (\ref{htp})  to
derive some physical quantities. Indeed,
from the standard definition, we show that
the particle number is
\begin{equation}\label{np}
    N^{\sf nc}_{\sf hi}=-N_{\phi}\left[2\alpha^{2} {\lambda\over\kappa} Li_{1}
    \left(-e^{\bar{\mu}/\lambda}\right)-
\left(\alpha ^{2}{g^\ast}^{2}-\frac{1}{3}\right)
    \frac{\ka}{16\lam\cosh^{2}\left({\bar{\mu}\over 2\lambda}\right)}\right]
\end{equation}
where in general for any integer $n$, the logarithmic function is given by
\beq
Li_{n}
    \left(-e^{\bar{\mu}\over\lambda}\right)=\sum_{l=1}^{\infty}
{1\over l^{n}}\ {\left(-e^{\bar{\mu}\over\lambda}\right)^l}.
\eeq
According to (\ref{edf}), the corresponding Hall conductivity
can be written as
\begin{equation}\label{hcht}
    \si^{\sf nc}_{\sf hi}= {e^2\over h}\left[2\alpha^{2} {\lambda\over\kappa} Li_{1}
    \left(-e^{\bar{\mu}/\lambda}\right)-
\left(\alpha ^{2}{g^\ast}^{2}-\frac{1}{3}\right)
    \frac{\ka}{16\lam\cosh^{2}\left({\bar{\mu}\over 2\lambda}\right)}\right]
\eeq
which is the general form that  can be derived
for particles of spin ${1\over 2}$ living on $\mathbb{R}^2_{\te}$.
One may study its behaviour at different values of $\te$ to
see what make difference with respect to low
temperature case.

At this level, let us discuss the magnetism that can be exhibited by
our  system at high temperature. Indeed, we  start by determining
its magnetization, which is
\begin{eqnarray}\lb{hmag}
\lefteqn{
M^{\sf nc}_{\sf hi} =- mc^2\lam {N_{\phi}\over B
  \kappa}\Biggl[4\lam\al\left(\al-1\right) Li_2\left(-e^ {\bar\mu/
    \lam}\right)-\left( \al{g^\star}^2 \left(2\al -1\right)
  -{1\over 3}\right)
{\kappa^2\over 2\lam\left(1+e^{-\bar{\mu}/\lam}\right)}~{} } \nonumber \\
& &
~~~
+{aE\over 2B}\left( 2\al^2 Li_1\left(-e^ {\bar\mu/
    \lam}\right) + \left(\al^2 {g^\star}^2-{ 1\over 3}\right)
\left({\ka\over 4\lam}\right)^2{1\over \cosh^2({\bar\mu\over
    2\lam})}\right)
\Biggr].
\end{eqnarray}
Therefore, we end up with the following consequences. From (\ref{htp}) and
(\ref{hmag}), it follows that  there is a phase transition, which is governed
by the critical point
\beq\lb{cpt}
g^* = {1\over \al\sqrt{3}}.
\eeq
More precisely,
 this is showing that the system is behaving like
a diamagnetic for $g^* < {1\over \al\sqrt{3}}$. However, it has
a paramagnetic nature if $g^* > {1\over \al\sqrt{3}}$.
This is interesting in sense that one may differently fix $\al$
to get different critical values and therefore different transitions
in real physical systems.

On the other hand, (\ref{cpt}) can be employed to experimentally
measure $\te$. In doing so, let us solve
 (\ref{cpt}) to find
\beq
\te =2l_B^2 \left( 1-{1\over g^*\sqrt{3}} \right).
\eeq
Using the experiment results given by (\ref{edata}), we obtain
\beq
\te = -36, 53 \ l_B^2
\eeq
which is showing that the phase of our system is changing
at this point.
Therefore, it gives
 another way of making use of the evidently experiment of
 the non-commutativity parameter.

One can also define a TPO density at high temperature limit. After calculation,
we end up with a new function as
\begin{eqnarray}\lb{hchi}
\lefteqn{
\chi_{\sf hi}^{\sf nc}=\left(1-\al\right) mc^2\lam {N_{\phi}\over B
  \kappa}\Bigg[4\lam\al Li_2\left(-e^ {\bar\mu\over
    \lam}\right)-{\al {g^\star}^2\kappa^2\over 2\lam
    \left(1+e^-{\bar{\mu}\over\lam}\right)}~{} } \nonumber \\
& &
~~~
-{a\over 2}\left(2\al^2 Li_1\left(-e^ {\bar\mu\over
    \lam}\right)-(\al^2 {g^\star}^2-{ 1\over 3})
\left({\kappa\over 4\lam}\right)^2{1\over \cosh^2({\bar\mu\over
    2\lam})}\right)\left({E\over B}\right)^2 \Bigg].
\end{eqnarray}
With this we complete our analysis as concerning the high temperature
regime.
Of course, some questions related to the Hall conductivity that
should be investigated deeply.

\section{Critical point}

Finally, let us consider the critical point corresponding
to the Hamiltonian describing  the present system on the plane $\mathbb{R}^2$.
In fact, we show that the Laughlin theory can be recovered
at this point. In doing so, let us return to
 (\ref{alpha}) in order to notice
\beq
\te_{\sf c} = 2l_B^2.
\eeq
Inserting this in the
non-commutative version of (\ref{HAM1}), we find a new Hamiltonian
\beq\lb{cham}
H^{\sf nc}|_{\te_{\sf c}} =  {{p}_y^2\over 2m} +
{m\om_c^2\over 2} y^2 + eE\left(y + {l_B^2\over \hbar} p_x\right).
\eeq
It can be simplified by returning to (\ref{nccoo})in order to
write down a new commutator as
 \beq [x, y ] = 2il_B^2.
 \eeq
Remember that this can be realized by dealing with the variable
$y$ as a canonical momentum of $ x$ and vis versa. This processing
leads to define
\beq y = 2{ l_B^2\over \hbar} p_x, \qquad x =- 2{
l_B^2\over \hbar} p_y.
\eeq
They allow us to write (\ref{cham}) as
\beq\lb{chams}
H^{\sf nc}|_{\te_{\sf c}} = {{p}_y^2\over 2m} +
{m\om_c^2\over 2} y^2 + {3\over 2}eE y.
\eeq

To recover the well-know results on QHE, one may handle
(\ref{chams}). This can be done by rearranging it to get
one-dimensional harmonic oscillator. Indeed,
 we simply make the following change
\beq
Y = y + {3eE\over m\om_c^2}
\eeq
 to end up with the
Hamiltonian
\beq\lb{cham1d} H^{\sf nc}|_{\te_{\sf c}} =
{{p}_Y^2\over 2m} + {m\om_c^2\over 2} Y^2  - {9\over 8} mc^2
\left({E\over B}\right)^2.
\eeq
This is actually describing the
lowest Landau level (LLL) of our system. Therefore, another theory
of QHE can be reformulated by analyzing  (\ref{cham1d}). Indeed,
it gives the Landau levels for the system confined in LLL, since
the term ${9\over 8} mc^2 \left({E\over B}\right)^2$ can be
dropped for a strong magnetic field. Thus, one may write $H^{\sf
nc}|_{\te_{\sf c}}$ as
\beq\lb{chamxy}
 H^{\sf nc}|_{\te_{\sf c}} =
{m\om_c^2\over 2}\left(X^2+ Y^2\right). \eeq
This is just a
confining potential and therefore  can be used to make contact
with the Laughlin theory~\cite{laughlin} for FQHE. Indeed, let us
define the creation and annihilation operators as
\beq Z = {1\over
2}\left(X-iY\right), \qquad \bZ={1\over 2}\left(X+iY\right)
\eeq
which satisfy the commutation relation
\beq
\left[Z, \bZ\right]=
\mathbb{I}.
\eeq
It is easily seen that $H^{\sf nc}|_{\te_{\sf
c}}$ can be mapped in terms of $Z$ and $\bZ$ to get
\beq\lb{chamz}
H^{\sf nc}|_{\te_{\sf c}}=  {\hbar\om_c\over 2}\left(Z\bZ+ \bZ
Z\right).
\eeq
Therefore, the corresponding spectrum is
\beq En
={\hbar\om_c\over 2} (2n+1), \qquad |n\ra= {Z^n\over \sqrt{n!}} |
0\ra, \qquad n\in {\mathbb N}.
\eeq

Now let us consider the picture of $N$-electrons in LLL. Its total Hamiltonian
is just sum of a single particle (\ref{chamz}), namely
\beq\lb{tot}
H_{\sf tot} =  {m\om_c^2\over 2}\sum_{i=1}^{N}
\left(X_i^2+ Y_i^2\right).
\eeq
It is obvious that its eigenvalues are just $N$ copies
of $E_n$. The corresponding
 many-body state can be written in
terms of the Slater determinant, such as
\begin{equation}\lb{lw2}
|1\rangle= \left\{\epsilon^{i_1 \cdots i_N} Z_{i_1}^{n_1} \cdots Z_{i_N}^{n_N}\right\}
|0\rangle
\end{equation}
which
is nothing but the first Laughlin state that corresponds to $\nu=1$~\cite{laughlin}.
Other similar Laughlin states can also be constructed
\begin{equation}\lb{lw3}
|m\rangle= \left\{\epsilon^{i_1 \cdots i_N} Z_{i_1}^{n_1} \cdots Z_{i_N}^{n_N}\right\}^m
|0\rangle
\end{equation}
corresponding to the filling factor $\nu ={1\over m}$, with $m$
odd integer. These shown that the critical point leads to an
interesting results and therefore needed to be investigated.

\section{Conclusion}

By considering a system of particles of spin ${1\over 2}$ living
 on the non-commutative plane in the presence of a magnetic and a
weak electric fields,
we have investigated its basic features from quantum statistical
 mechanical
point of view. After writing down the corresponding Hamiltonian,
we have derived its energy levels as well as the wavefunctions.
They have been used to analyze the thermodynamical properties of
our system. More precisely, we have given
 the partition function that allowed us to
get the associated thermodynamical potential (TPO). Since a
general form of TPO have been derived, we have introduced the
Mellin transformation to get a simplified TPO. To completely
evaluate this latter, we have introduced some condition
supplemented by separating two different cases: low and high
temperatures.

In the first case, we have derived a general form of
the Hall conductivity that lead to different results.
Indeed, from our consideration and making using of the
experiment data, we have given a way to measure the
non-commutativity parameter $\te$. Subsequently, this has
been tuned on to make contact with different results
for QHE. In particular, we have obtained integer QHE simply
by switching off $\te$. On the other hand, the
observed fraction values, characterizing fractional QHE,
have been recovered
by fixing $\te$.
Moreover, we have shown that
 $\te$ can be chosen to interpret our
system as a collection of composite fermions or multilayer Hall
 systems on the ordinary plane.



In the second case, we have derived the particle number as well as
the corresponding Hall conductivity. To analyze the magnetism of
our system, we have  evaluated the magnetization. This allowed us
to get  a critical value in terms of $\te$. This has been used to
discuss the phase transition between two sectors. In fact, they
are obtained to be diamagnetic and paramagnetic of nature.
Furthermore, at the critical point, we have used the experiment
data to fix $\te$.

 Finally we have analyzed the critical point of our
non-commutative Hamiltonian, i.e. $\te=2l_B^2$. By making use of
some rearrangement, we have ended up with one-dimensional harmonic
oscillator. This latter is sharing many features with that
corresponding to the lowest Landau level. Moreover, it has been
used to make contact with the Laughlin theory.

Still some questions to be addressed. At the critical point,
one may think about making
use the matrix model theory to investigate
thermodynamical properties of the system in LLL. On the other
hand, a link to the spin Hall effect deserves an attention.
We hope to return to these issues and related matter
in forthcoming works.



\section*{Acknowledgments}

The authors are thankful to M. El Bouziani for
fruitful discussions on the phase transition.
AJ~work's was partially
supported by Arab Regional Fellows Program (ARFP) $2006/2007$.


\begin{thebibliography}{1}

\bibitem{prange}
R.E. Prange and S.M. Girvin (editors), {\it The Quantum Hall
Effect}, (New York, Springer 1990).

\bibitem{kli} K. von Klitzing, G. Dorda, M. Pepper, {\it
  Phys. Rev. Lett.}
{\bf 45} (1980) 494.

\bibitem{connes} A. Connes, {\it Noncommutative geometry}, (Academic
  press, London 1944).


\bibitem{sakita} S. Iso, D. Karabali and B. Sakita,  {\it Nucl.Phys.}
{\bf B388} (1992) 700,  {\sf hep-th/9202012}.

\bibitem{jelnpb} A. Jellal, \NP {\bf B725} (2005) 554,  {\sf hep-th/0505095}.

\bibitem{landau} L.D. Landau and E.M. Lifshitz, {\it Quantum Mechanics
$3^{rd}$ edition}, (Pergamon, London, 1977).

\bibitem{hky} C-L. Ho, V.R. Khalilov and C. Yang, {\sf cond-mat/9606020}.


\bibitem{dj} \"O.F. Dayi and A. Jellal, \JMP {\bf 43} (2002) 4592,
{\sf hep-th/0111267}.

\bibitem{ich} Y. Ishikawa and H. Fukuyama,
{\it J. Phys. Soc. Jap.} {\bf 68} (1999) 2405,  {\sf cond-mat/9904052}.

\bibitem{gaz} J.P. Gazeau, P.Y. Hsiao and A. Jellal,
\PR {\bf B65}  (2002) 094427,  {\sf cond-mat/0101338}.

\bibitem{jel1} A. Jellal,  \JP A  {\bf 34} (2001) 10159,
{\sf hep-th/0105303}.

\bibitem{stone} H.L. St\"ormer, {\it in Frontiers in Physics,
{High Technology and Mathematics}},
H.A. Cerdeira and S.O. Lundqvist (editors), (World Scientific, Singapore, 1990).


\bibitem{qhe} D.C. Tsui, H.L. St\"ormer and A.C. Gossard,
\PRL {\bf 48} (1982) 1555; one may see also reference  $[1]$.

\bibitem{cf} J.V. Jain, \PRL {\bf 63} (1989) 199; \PR {\bf B41}
(1990) 7653; {\it Adv. Phys.} {\bf 41} (1992) 105;
O. Heinonen (editor), {\it Composite Fermions: A Unified View of Quantum Hall
Regime}, (World Scientific, 1998).

\bibitem {tsui} T. Haavasoja, H.L. St\"ormer, D.J. Bishop, V. Narayanamurti,
 A.C. Gossard and W. Wiegmann, {\it Surf. Sci.} {\bf 142} (1984) 294.

\bibitem{manar} A. Jellal, \MPL {\bf A18} (2003) 1473, {\sf hep-th/0209259}.

\bibitem{laughlin}R.B.~Laughlin, \PRL
{\bf 50} (1983) 1395.


\end{thebibliography}
\end{document}